\documentstyle[seceq]{ptptex}

\newcommand{\be}{\begin{equation}}
\newcommand{\ee}{\end{equation}}
\newcommand{\bea}{\begin{eqnarray}}
\newcommand{\eea}{\end{eqnarray}}

\newcommand{\ri}{\mbox{i}}
\newcommand{\re}{\mbox{e}}

\title{Chiral order in spin-S XY chains}

\author{
Thierry {\sc Jolic\oe ur}\footnote{e-mail~:
Thierry.Jolicoeur@lpmc.ens.fr} and Philippe {\sc
Lecheminant}\footnote{A common friend of Fermi and Bose. e-mail~:
phle@lptl.jussieu.fr} }

\inst{$^*$Laboratoire de Physique de la Mati\`ere Condens\'ee,
Ecole Normale Sup\'erieure, 24 rue Lhomond, 75231 Paris, France
\\
$^{**}$Laboratoire de Physique Th\'eorique et Mod\'elisation,\\
Universit\'e de Cergy Pontoise, 5 Mail Gay-Lussac, Neuville sur
Oise, 95301 Cergy Pontoise, France}

\recdate{\today }
\abst{ We consider the issue of chiral ordering in spin chains for
an arbitrary value of the spins S. By use of bosonization
according to a scheme developped by H.-J. Schulz, we obtain the
phase diagram of the chain with XY exchange couplings between
nearest and next-to-nearest neighbors. We obtain a satisfactory
picture  including a so-called chiral spin nematic phase which is
gapless, has long-range chiral order and incommensurate spin
correlations. We perform a stability analysis of this phase and
point out that this analysis is in conflict with existing DMRG
results that shows a difference between integer and half-integer
spin case.
 }

\begin{document}

\maketitle

\section{Introduction}
Quantum antiferromagnetic spin chains display a variety of phases
that have no classical counterpart. This variety is even increased
if we consider the effect of frustration. Many studies\cite{J1J2}
have been devoted in the past to the simple AF chain with nearest
neighbor exchange $J_1$ and next-to-nearest neighbor $J_2$. In the
isotropic case, the Hamiltonian is then simply~:
\be
{\cal H} =  J_1 \sum_n\left({\bf S}_n {\bf S}_{n+1} \right) + J_2
\sum_n\left({\bf S}_n {\bf S}_{n+2} \right),
\label{heisenbergfrus} \ee where $S_n^{\pm} = S_n^x \pm i S_n^y$
is a spin operator at site n and there are competing
antiferromagnetic interactions $J_1,J_2 > 0$ which introduces
frustration in the model. In the spin-1/2 case, for small $J_2$,
there is a spin-fluid phase whose effective theory is that of a
massless free boson. It has quasi-long range spin order with
algebraic decay of spîn correlations. For larger values of $J_2$,
the ground state is spontaneously dimerized~: this appears through
a quantum phase transition of Kosterlitz-Thouless type and
eventually incommensurability develops within this gapped phase
for even larger values of $J_2$ but without any additional phase
transition.The situation is remarkably different if we now
consider XY exchange~:
\be
{\cal H} =  J_1 \sum_n\left(S^x_n S^x_{n+1} + S^y_n
S^y_{n+1}\right) + J_2 \sum_n\left(S^x_n S^x_{n+2} + S^y_n
S^y_{n+2}\right). \label{hamxyfrus} \ee Starting from the limit
with large $J_2$, Nersesyan et al.\cite{nersesyan} used a
mean-field treatment and then bosonization to predict the
occurence of a new phase with many unconventional characteristics.
In this limit, they predicted that there is long-range {\it
chiral} order~:
\be
\langle ({\vec S}_n \wedge {\vec S}_{n+1})_z \rangle \ne 0.
\label{chiralorder} \ee There are also local spin currents
polarized along the anisotropy $z$-axis. This phase is gapless and
there are incommensurate spin correlations that decay
algebraically with an exponent which they found to be 1/4. The
existence of this phase has been recently demonstrated
numerically. Such a phase has been disclosed in a ladder model
formulated first as an array of Josephson junctions which is
equivalent to a spin-1/2 half model\cite{nishiyama}. this ladder
has square plaquettes so it is not frustrated as model
Eq.(\ref{hamxyfrus}) but frustration is introduced by half a flux
quantum piercing the plaquettes. Then a study\cite{HKK2000} using
the DMRG algorithm gave evidence for this phase in model
Eq.(\ref{hamxyfrus})~: the spin fluid phase is stable up to
$J_1/J_2\approx 0.33$ then the chain undergoes dimerization and at
$J_1/J_2\approx 1.26$ there is a second transition to the chiral
critical phase. This new phase has also been
reported\cite{KKH99,HKKT2000} in the S=1 chain with the same
Hamiltonian Eq.(\ref{hamxyfrus}). Here the situation is even more
richer. When $J_2=0$, we are in the XY phase which destroyed
immediately by adding even an infinitesimal $J_2$, the phase that
appears then is the celebrated gapped Haldane phase. This phase
resists the perturbing influence of $J_2$ for a while but at
$J_1/J_2\approx 0.47$ there is a phase transition to chiral order
but the gap remains nonzero. Then very close, at $J_1/J_2\approx
0.49$, there is a {\it distinct} transition to a critical phase
with chiral order as in the S=1/2 case. So the integer S=1 case
has an additional phase w.r.t. the S=1/2 case, a gapped chiral
phase. This difference persists for higher spins\cite{HKK2001}.
For S=3/2, there is an intermediate dimerized phase which is
replaced in a single tranistion by by chiral critical phase. For
S=2, the Haldane phase is destroyed by two successive transitions
as  for S=1. To understand this pattern of phase transition we
use\cite{PJ} the bosonization technique which has been adapted to
the case of generic spin-S by Schulz\cite{schulz}. This method is
able to capture the phase diagrams as a function of exchange
anisotropy as well as single-ion anisotropy and it correctly
captures the difference between integer and half-integer spins.

\section{Weak coupling limit}
The idea is to write down each spin-S as a sum spins 1/2~:
\be
S^{\pm}_n  = \sum_{a=1}^{2S} s^{\pm}_{a,n}. \label{sumspins} \ee
Each spin 1/2 is then bosonized by the standard technique. There
is a bosonic field $\phi_a, a=1\dots 2S$ associated with each
spin. The first possibility is to treat $J_2$ as a perturbation.
So we start from an isolated spin-S chain and write its effective
theory in terms of the bosons $\phi_a$. Due to the appearance of
couplings $s_a^+ s_b^-$, there are operators that induce gaps for
some linear combination of the basic bose field. More precisely,
only the "acoustic" mode remains massless~:
 \be
 \Phi
=\frac{1}{\sqrt{2S}} \left(\varphi_{1} + ... + \varphi_{2S}
\right) . \label{acoustic} \ee The effective Hamiltonian for the
acoustic mode is thus a simple free theory~:
\be
{\cal H}_{XY1} \simeq \frac{v}{2}\left( \Pi^2 +\left(\partial_x
\Phi\right)^2  \right), \label{acousticHam} \ee where $\Pi$ is the
canonical momentum conjugate to $\Phi$. The coefficient $v$ is an
unimportant velocity and we have used the conventional name "XY1"
coming from the standard S=1 chain phase diagram. The spin
operator can be expressed in terms of the field $\Theta$ which is
{\it dual } to $\Phi$~:
\be
S^{\pm} \sim \left(-1\right)^{x/a} \exp\left(\pm \ri
\sqrt{\pi/2S}\; \;\Theta\right). \label{spinOp} \ee This
expression shows easily that the XY spin correlations decay
algebraically with an exponent $\eta=1/4S$. This is in agrrement
with numerical findings\cite{alcaraz}. In the S=1 case it is also
in agreement with work by Kitazawa et al.\cite{kitazawa}. If we
now bosonize the perturbation $J_2$, we find that there is a
simple renormalization of the previous free hamiltonian
Eq.(\ref{acousticHam}) but in addition vertex operators appear in
perturbation expansion. For integer spin, the most relevant
operator appear at S$^{th}$ order and it appears at 2S$^{th}$
order for half-integer spins. The effective theory is then~: \be
{\cal H} \simeq \frac{v}{2}\left(K \Pi^2 + \frac{1}{K}
\left(\partial_x \Phi\right)^2 \right) - \frac{g_{eff}}{a}
\cos\left(\beta \Phi\right), \label{J2effect} \ee where
$\beta=\sqrt{8\pi S}$ for integer spins and $\beta=\sqrt{32\pi S}$
for half-integer spins. The Luttinger parameter K is obtained in
perturbation $K=1-(4/\pi) J_2/J_1 +O(J_2^2)$. This effective leads
then immediately to the phase diagram of the spin-S XY chain for
small $J_2/J_1$. The scaling dimension of the vertex operator in
the effective theory Eq.(\ref{J2effect}) is $K\beta^2/4\pi$ and
thus is irrelevant for small $J_2$, the XY1 phase will have thus a
finite extent. With increasing $J_2$, the vertex operator becomes
relevant and drives the system towards a massive phase through a
KT transition. Depending upon the spin parity, it will be the
Haldane or the dimerized phase. If we take seriously the
approximate formula for $K$, we deduce $J_2/J_1\approx 0.29$ at
the KT transition for S=1/2, which compares quite favorably to the
numerical estimate of $\approx 0.324$. We also predict that S=1 is
special~: in this case indeed the operator is marginal for
$J_2\rightarrow 0$ hence the instability takes place immediately
upon switching an infinitesimal value of $J_2$, as seen in DMRG
studies.

\section{Zigzag limit}
We now turn to the opposite limit $J_2 \gg J_1$. Then we have a
two-leg spin-S XY ladder coupled in a zigzag way. We first
bosonize the two independent legs when $J_1=0$. So each of the
chains can be treated as in the previous section. There are now
two acoustic modes $\Phi_1$ and $\Phi_2$ that are the effective
low-energy degrees of freedom. It is convenient to introduce the
symmetric and antisymmetric combinations of these two modes~:
\be
{ \Phi}_{\pm} = \frac{1}{\sqrt{2}}\left(\Phi_{1} \pm
\Phi_{2}\right). \label{combis} \ee The leading contribution to
spin correlations comes from the following operator~:
\be
S_a^{\pm} \simeq \frac{\lambda}{\sqrt{a}} \left(-1\right)^{x/a}
\exp\left(\pm \ri \sqrt{\pi/2S}\; \;\Theta_{a}\right),
\label{spinOps} \ee where $a=1,2$ and $\Theta_a$ are the fields
dual to $\Phi_a$. When $J_1=0$ the two fields $\Phi_a$ are free
and massless. Introducing $J_1$, we find the effective theory~:
\be {\cal H} \simeq \frac{v}{2}\sum_{a=\pm}\left(\Pi^2_a+\left(
\partial_x { \Phi}_a\right)^2
\right) + g \;
\partial_x { \Theta}_+ \sin\left(\sqrt{\frac{\pi}{S}}{
\Theta}_-\right), \label{J1effect} \ee where $\Theta_\pm$ are
fields dual to $\Phi_\pm$ and $\Pi_\pm$ are canonical conjugate to
$\Phi_\pm$, and $g=O(J_1)$. The operator perturbing the free part
in Eq.(\ref{J1effect}) is a parity symmetry breaking term with
nonzero conformal spin (=1). Its effect is thus highly nontrivial.
A simple perturbation with nonzero conformal spin is given by the
uniform component of the spin density $\partial_x\Phi$. In this
case we know its effect~: it induces incommensurability. We treat
theory Eq.(\ref{J1effect}) by following exactly the method of
nersesyan et al. We just decouple~:
\be
\partial_x { \Theta}_+ \sin\left(\sqrt{\frac{\pi}{S}}{
\Theta}_-\right)\rightarrow \kappa <\partial_x {
\Theta}_+>+\mu<\sin\left(\sqrt{\frac{\pi}{S}}{ \Theta}_-\right)>.
\label{decouple} \ee We then impose self-consistency. The "+"
sector remains massless while the "-" sector is massive due to the
vertex operator in Eq.(\ref{decouple}). The results that follow
are then close to the original findings for for S=1/2 critical
chiral phase. The most notable difference is that spin
correlations decay algebraically with a spin-dependent exponent~:
\be
\langle S_1^{\dagger}\left(x\right) S^{-}_a\left(0\right) \rangle
\sim \frac{\re^{iq x}}{|x|^{1/(8S)}}, \; \; a=1,2 , \label{Sdecay}
\ee with $q-\pi/a \sim (J_1/J_2)^{4S/(4S-1)}$. The exponent 1/4 of
the S=1/2 case is thus a special case of $\eta=1/8S$. This formula
is in good agreement\cite{HKK2001} with measurements by DMRG up to
S=2. There are nontrivial spin currents in the ground state~:
\begin{equation}
\langle J_{1s}^{z} \rangle = \langle J_{2s}^{z} \rangle = -v
\sqrt{\frac{S}{\pi}} \; \langle \partial_x { \Theta}_{+} \rangle
\ne 0 . \label{spincurrent}
\end{equation}
In the language of spins, this means intrachain currents~:
\begin{equation}
\langle \left({\vec S}_{a,n} \wedge {\vec S}_{a,n+1}\right)_z
\rangle \propto  \sqrt{\frac{\pi}{4S}}  \; \langle
\partial_x { \Theta}_{+} \rangle \ne 0, \; \; a=1,2,
\label{chiralorderinchain}
\end{equation}
as well as interchain currents~:
\begin{equation}
J_1 \langle \left({\vec S}_{1,n} \wedge {\vec S}_{2,n}\right)_z
\rangle \propto  \langle \sin\left(\sqrt{\frac{\pi}{S}} {
\Theta}_- \right) \rangle \ne 0, \label{chiralorderinterchain}
\end{equation}
this also means long-range {\it chiral order}.

\section{Stability analysis}
We certainly expect that when increasing the coupling $J_1$ the
"+" sector of theory (\ref{J1effect}) will not remain massless
forever because we know from the analysis of the previous approach
in the small $J_2$ limit that there is a massive phase. It is
likely that some vertex operator will become relevant at some
finite value of $J_1/J_2$. We try to find this operator by
analyzing the symmetry properties of the theory. Taking into
account the symmetries of the {\it mean-field} Hamiltonian, we
find the following effective theory~: \be {\cal H}_+ \simeq
\frac{v}{2}\left(K \Pi_+^2 + \frac{1}{K} \left(\partial_x {
\Phi}_+\right)^2 \right) + \kappa
\partial_x { \Theta}_+ - \frac{g_{eff}}{a} \cos\left(\gamma { \Phi}_+\right),
\label{guessStability} \ee where $\gamma=\sqrt{16\pi S}$ for
integer S and $\gamma=\sqrt{64\pi S}$ for half-integer spins. When
this vertex operator becomes relevant, the system is gapped and
still incommensurate. This is consistent with the observation of a
"chiral Haldane" phase in numerical studies for integer spins. For
half-integer spin, this predicts a phase which is dimerized (due
to the value at which $\Phi_+$ is pinned) and has chiral LRO. We
are then forced to speculate that there is then an additional
Ising transition at which chiral LRO disappear to make contact
with the small $J_2$ limit.

\section{Conclusions}
The bosonization approach is able to reproduce the phase diagram
of the XY $J_1-J_2$ chain for integer spins. For all spins it
correctly predicts the existence of the chiral critical phase. The
spin correlations decay with exponent $1/8S$. For the half-integer
case, this approach predicts a dimerized phase with
incommensurability and chiral order which is apparently not seen.
It is possible that this signals a failure of the mean-field
decoupling. For integer spins, there is good agreement with a
large-S study\cite{Alexei}. A possible candidate\cite{Kikuchi} for
the chiral critical phase for S=1 is CaV$_2$O$_4$ which has
$J_1\approx J_2$. This would require a large single-ion
anisotropy\cite{H2001} to escape from the double Haldane phase.

\section*{Acknowledgements}
We would like to thank T. Hikihara and M. Kikuchi for information
on their recent results.

\end{document}